# Investigation of chemical structure recognition by encoder-decoder models in learning progress

Shumpei Nemoto[1, *]   Tadahaya Mizuno[2, *, †]   Hiroyuki Kusuhara[1]

[1] Laboratory of Molecular Pharmacokinetics, Graduate School of Pharmaceutical Sciences, The University of Tokyo, 7-3-1 Hongo, Bunkyo, Tokyo, Japan
[2] Laboratory of Molecular Pharmacokinetics, Graduate School of Pharmaceutical Sciences, The University of Tokyo, 7-3-1 Hongo, Bunkyo, Tokyo, Japan, tadahaya@gmail.com
[†]Author to whom correspondence should be addressed.
[*]These authors contributed equally.

## Abstract

Descriptor generation methods using latent representations of encoder-decoder (ED) models with SMILES as input have been actively developed in recent years. However, it is not clear how the structure is recognized in the learning progress of ED models. In this work, we created ED models of various learning progress and investigated the relationship between structural information and learning progress. We showed that compound substructures were learned early in ED models by monitoring the accuracy of downstream tasks and input–output substructure similarity using substructure-based descriptors, which suggests that existing evaluation methods based on the accuracy of downstream tasks may not be sensitive enough to evaluate the performance of ED models with SMILES as descriptor generation methods. On the other hand, we showed that structure restoration was time-consuming, and in particular, insufficient learning led to the estimation of a larger structure than the actual one. It can be inferred that determining the endpoint of the structure is a difficult task for the model. To our knowledge, this is the first study to link the learning progress of SMILES by ED model to chemical structures for a wide range of chemicals.

**Keywords**: Descriptor, Encoder-Decoder model, learning progress, chemical structure, GRU

## 1 Introduction

The numerical representation of chemicals is useful for grasping their properties and is roughly divided into two approaches: phenotype-based and structure-based representation.[1]–[3] The latter is often called a descriptor, and determining how to generate a *good* descriptor is one of the big topics in the field of chemoinformatics.[4], [5] Since the report by Gómez-Bombarelli et al. in 2016, descriptor generation methods using latent representations of encoder–decoder (ED) models with Simplified Molecular Input Line

Entry Specification (SMILES) as input have attracted much attention.[6] The ED model in natural language processing (NLP) is a model that encodes strings into numerical information once and decodes them into strings again.[7], [8] The numeric information has rich information describing the strings and is called the latent representation.[9] Therefore, numerical information of chemical structures (descriptors) can be obtained by ED models learning SMILES, with the string representation of chemical structures as the input. In particular, the idea of neural machine translation (NMT) was introduced by Winter et al. in 2019.[10] For example, to translate Japanese to English correctly, the ED model must understand not only the characters of both Japanese and English but also the context of the strings.[11], [12] The latent representation of NMT models learning SMILES includes the context of SMILES, i.e., the entire chemical structure.[13]–[15]

Descriptor generation methods based on ED models with SMILES as input have two characteristics: they provide continuous descriptors and can transform numerical information back into structures. Most conventional descriptors in the field of chemoinformatics, such as Extended Connectivity Fingerprint (ECFP), are binary vectors based on substructures, restricting the descriptive ability of structures and restoration to the original structures.[16], [17] The machine learning-based methods before ED models, such as mol2vec and Neural FingerPrint (NFP), are continuous representations and relatively high structure representability, but none of them can restore chemical structures from descriptors.[18], [19] Because of their ability to convert numerical information into a structure, ED models, together with other generative models such as generative adversarial networks (GANs),[20] are used in de novo drug design.[21]–[24] In contrast, models in the context of de novo drug design are often end-to-end models and do not cover as wide a variety of chemical structures as descriptor generation methods do.[25]

Many descriptor generation methods based on ED models with SMILES as input have been developed.[6], [10], [26] As with other descriptor generation methods, many of these studies discuss the merits of the methods based on the accuracy of some prediction model (e.g., mode of action prediction) using the generated descriptors based on the idea that "*Good representation leads to good downstream results.*" In addition, the structure restorability has not been examined well and is only ensured by confirming the input–output consistency is higher than a certain level. Thus, descriptor generation methods based on ED models have been investigated in an indirect fashion with chemical structures. The relationship between the ED models fed with SMILES for descriptor generation (not for end-to-end tasks) and chemical structures, such as how the ED models learn and recognize the structures of various chemicals, is currently unknown despite the widespread use of models with SMILES.

The purpose of this study is to clarify how the structures of various compounds are recognized in the learning progress of ED models for descriptor generation. *Recognition* of chemical structures by the ED model is defined here as the ability to obtain numerical information reflecting the chemical structure and to reconstruct the chemical structure from the numerical information (structural representation and restoration). We created a model set consisting of ED models with various learning progress and investigated the relationship between learning progress and chemical structures.

## 2 Method

**Data preparation**

The chemical data set containing SMILES representations was obtained via ZINC15,[27] and 30 million chemicals were randomly extracted for training the ED model. The following criteria were used to filter the chemicals inspired by Le et al.[16] (1) only containing organic atom set, (2) The number of heavy atoms between 3 and 50. Note that these filtrations were employed simply to facilitate modeling, but can introduce bias into the ED model results (an investigation into their impact is beyond the scope of this study). The salts were stripped and only the largest fragments kept. A random SMILES variant was generated using the SMILES enumeration procedure.[28] For the evaluation of the descriptor, the high throughput screening (HTS) assay data set was obtained via EPA.[29] The data were processed via an R library, ToxCast-tcpl.[30] The transcriptome profile data set of MCF7 was obtained via iLINCS.[31] The SMILES representations of chemicals in the transcriptome profile data were obtained from PubChem[32] using PubChemPy 1.0.4, PubChem API used in Python. In this study, we used RDKit (ver. 2022.03.02) for handling molecules such as SMILES generation.[33]

**Model preparation**

When creating the model, we referenced the model architecture developed by Winter et al.,[10] however, we modified the bucketing strategy to handle stereochemistry, which is removed in Winter's model (refer to the Availability section for details). The encoder network consists of the 3-layer gated recurrent unit (GRU) with 256, 512, and 1024 units, followed by a fully connected layer that maps the concatenated cell states of the GRU to the latent space with 256 neurons and hyperbolic tangent activation function. The decoder network takes the latent space as input and feeds it into a fully connected layer with 1792 neurons. This output is split into three parts and used to initialize 3-layer GRU cells. The complete model is trained on minimizing the cross-entropy between the output of the decoder network, the sequence of probability distributions over the different possible characters, and the one-hot encoded correct characters in the target sequence. For the decoder GRU, we utilized teacher forcing.[34] For a robust model, a 15% dropout was applied, and noise sampled from a zero-centered normal distribution with a standard deviation of 0.05 was added to concatenated cell states of the GRU in the encoder.

Thirty million chemicals extracted from the ZINC data set were applied as a training set for training the models. The model was trained on translating from random SMILES to canonical SMILES. The Adam optimizer was used with a learning rate of $5 \times 10^{-4}$, and an exponential scheduler was used that decreases the learning rate by a factor of 0.9 every 250 epochs. The batch size was set to 1024. To handle sequences of different lengths, we trained models for 6, 13, 104, 260, 338, and 598 epochs to obtain models of varying accuracy. We used the framework PyTorch 1.8.0 to build and execute our proposed model. The models were trained on 1 NVIDIA RTX3090 GPU. To evaluate the model accuracy, as an evaluation index, we defined perfect accuracy and partial accuracy, represented by the following equations:

$$\textbf{perfect accuracy} = \frac{1}{n}\sum_{i}^{n} I(t = p)$$

$$\textbf{partial accuracy} = \frac{1}{n}\sum_{i}^{n} \left\{ \frac{1}{\max(l(t), l(p))} \sum_{j}^{\min(l(t), l(p))} I(t_i = p_i) \right\}$$

where $n$ is the number of chemicals in the evaluation set, $t$ is the correct SMILES, $p$ is the predicted SMILES, $t_i$ is the ith letter of the correct SMILES, $l(t)$ is the length of $t$, and $I(x)$ is the function in which $I(x)$ is 1 if the prediction is correct and 0 otherwise.

The bottleneck layer of the ED model is regarded as the low-dimensional representation of chemicals with 256 dimensions and can be regarded as the descriptor. Chemical descriptors are obtained by feeding SMILES to the encoder of the trained model.

**Training HTS data**

ToxCast assay data were predicted with descriptors obtained from encoders as inputs. We selected XGBoost as a representative machine learning method, and hyperparameters were optimized using Optuna for each assay and threefold cross validation was applied.[35], [36] Optimized hyperparameters and the conditions for optimization are provided in **Supplementary Table 1**.

The HTS assay was filtered by the following criteria: (1) containing more than 7000 experimental chemicals, (2) the ratio of active chemicals to total is higher than 5%, and we extracted 113 assays (listed in **Supplementary Table 2**). 25% of each assay data was split for the test set and used to evaluate the trained model. We used two evaluation indexes of model accuracy, the area under the receiver operating characteristic curve (AUROC) and the Matthews correlation coefficient (MCC).

**Visualization of chemical space**

To visualize the distribution of chemicals formed by descriptors, dimensionality reduction was performed using UMAP,[37] and descriptors of 292 CMap chemicals were subjected to the algorithm. To understand the difference of chemical spaces obtained from models with different accuracies, the ECFP of each chemical and the Tanimoto coefficient of any two ECFPs were calculated.[38] Based on the Tanimoto coefficient, we defined three chemical groups with similar structures: (1) coefficient with estradiol is higher than 0.25 (estrogens) except for Fluvestrant because of the long hydrocarbon chain that is inappropriate for a similar structure with estradiol, (2) coefficient with apigenin is higher than 0.25 (flavonoids), (3) coefficient with isoconazole is higher than 0.25 (azoles). A scatter plot of chemicals embedded by UMAP was formed and chemicals in the three groups were highlighted. The chemicals in each group are listed in Supplementary Table 3.

**Investigation of substructure learning**

Using one of the 100K compounds set used for model evaluation, we extracted structures that generated valid structures in all models whose accuracy was higher than Model_0.1. MACCS Key and 2048-bit ECFP were calculated for the input and generated structures for each compound, and the agreement was evaluated using the Tanimoto coefficient.[17]

**Investigation of structure incorrectly decoded**

Using one of the 100K compounds set used for model evaluation, the string length and molecular weight of actual SMILES versus that of predicted SMILES that were not correctly decoded during the evaluation of each model were calculated. These are shown in a scatterplot, together with a straight line with $y = x$ (the value of actual SMILES is equal to that of predicted SMILES).

## 3 Results and Discussion

### 3.1 Preparation of Encoder-Decoder model with various accuracy: various ED model set

To investigate how the learning progress of ED models is related to the recognition of chemical structures, we first constructed a set of models with different learning progress. We randomly selected 30 million compounds from the ZINC database and prepared several models with different translation accuracies by controlling the learning progress.[27] As a negative control, we also prepared a set of 0-epoch models that were not trained at all and constructed a set of models together.

Five sets of 10K, 100K, and 1M compounds other than those used in the training were randomly selected from the ZINC database as well, and we evaluated the translation accuracy of each model on these sets. In addition to perfect accuracy, which evaluates the perfect agreement between input and output for easy interpretation as translation accuracy, partial accuracy, which evaluates partial recognition, was also employed as a measure of model accuracy (Figure 1a). The results showed that both accuracy indices increased as the training progressed, as expected. Notably, even when perfect accuracy was as low as 0.1% and 29%, partial accuracy was relatively high at 26% and 63%, respectively (Figure 1b). The presented results indicate that even when the whole string is not correctly restored, learning of substrings progresses at a relatively early stage.

In the following, the set of ED models with various translation accuracies will be referred to as the "various ED model set," and each model in the various ED model set will be denoted by its perfect accuracy. For example, the 598 epoch model showed 94% complete accuracy, so it will be denoted as "Model_94."

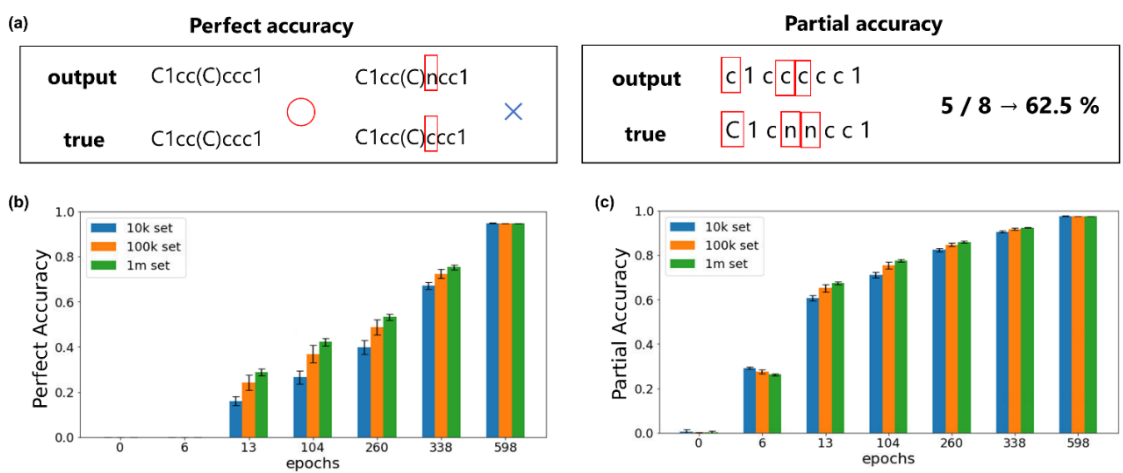

**Figure 1. Preparation of Encoder-Decoder model with various accuracy.** (a) Methodology of Encoder-decoder Model evaluation. (b) Comparison of evaluation metrics for evaluation dataset by each model.

### 3.2 Relationship between learning progress and downstream tasks with latent representation

performance of descriptor generation methods is evaluated based on the accuracy of downstream tasks using descriptors. To evaluate the relationship between the learning progress of ED models and their structure representability, we constructed models to estimate the properties of chemicals in ToxCast data sets using descriptors generated by each model in the various ED model sets and compared the accuracy of the property prediction. Figure 2a shows the AUROC and MCC[39] for 10 representative assays out of 113 assays derived from ToxCast (refer to the Methods section for details). The average of the AUROC of all 113 assays is shown in **Supplementary Figure 1**. Model_0 has an AUROC close to 0.5, indicating that these end-to-end tasks cannot be solved by the descriptors generated by the models without learning. On the other hand, the remaining models with learning all showed similar and relatively high prediction accuracy. Similar relationships with learning progress tasks were also confirmed in other datasets: two regression tasks (lipophilicity and solubility prediction) with Lipiphilicity and FreeSolv datasets (**Supplementary Figure 2**) and three classification tasks in MoleculeNet dataset (**Supplementary Figure 3**). The representative 10 assays of ToxCast HTS data were predicted using conventional descriptors, ECFP and

Mordred, with accuracy comparable to the most accurate ED model descriptors, which is consistent with previous reports (**Supplementary Figure 4**).

Next, we worked on the evaluation of structure representability in terms of chemical space. The connectivity map data set (292 compounds) was selected as a relatively small-size chemical set and visualized after dimensionality reduction with UMAP.[31], [37] To evaluate visually the closeness of chemical space of structurally similar chemicals, we prepared groups of similar chemical structures based on the Tanimoto coefficient, which is a similarity index of representative compound structures, and information on compound classes (Supplementary Table 3). As a result, in model_0, similar chemical structure groups were scattered, and no clear trend was observed. On the other hand, the remaining models with training showed that similar chemical structure groups were distributed in each specific region (Figure 2b).

These results suggest that among the properties of ED model-based descriptor generation methods, structure representability such as downstream task accuracy and chemical space are acquired early in the learning progress.

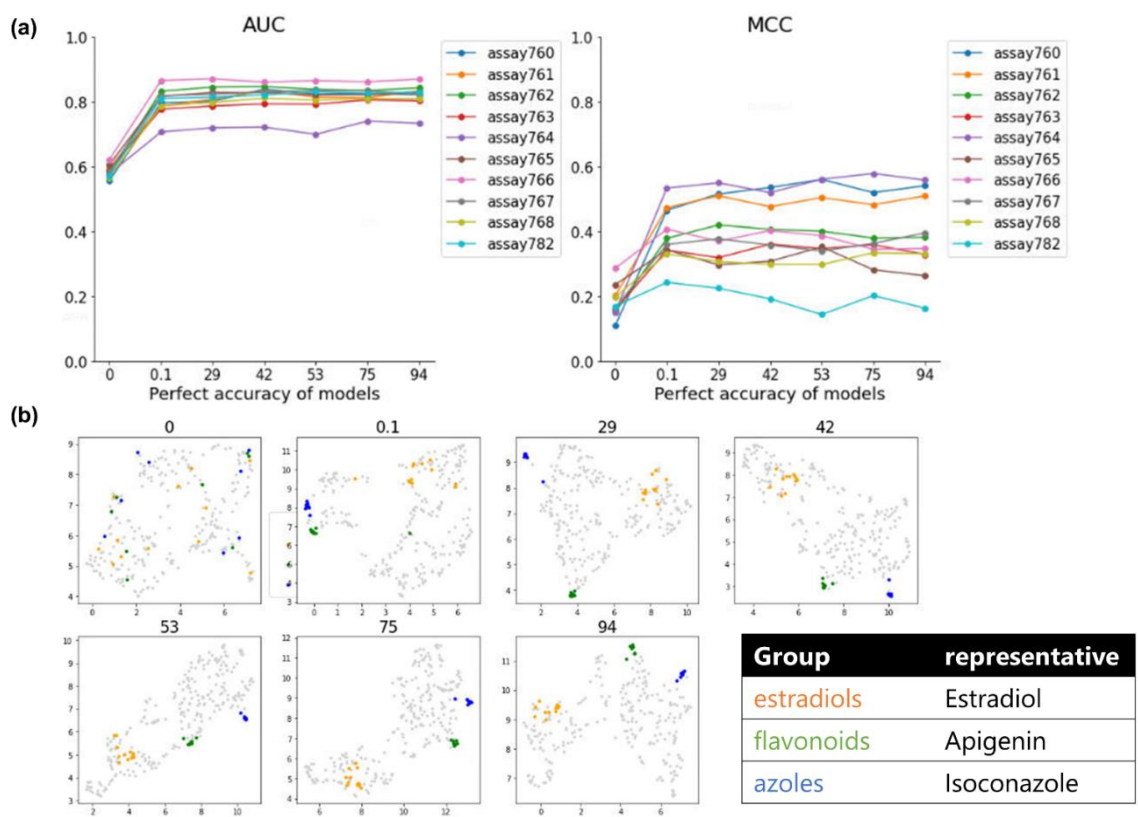

**Figure 2**. **Relationship between learning progress and downstream tasks with latent representation.** (a) AUROC and MCC of 10 representative assays prediction by XGBoost. (b) Scatter plots of 292 compounds obtained from CMap with dimension reduction by UMAP. Orange, green, and blue points indicate compounds with Tanimoto coefficients of 0.25 or greater with estradiol, apigenin and isoconazole, respectively.

### 3.3 Relationship between learning progress and substructure recognition

Considering the correlation between the partial accuracy in **Figure 1** and the accuracy of downstream tasks in **Figure 2** with respect to the learning progress, it is inferred that ED models recognize substructures of chemicals in the early stage of learning progress. Next, focusing on substructures, we evaluated the relationship between learning progress and structural restorability of the ED model.

Conventional descriptors such as MACCS and ECFP are discrete representations of binary vectors based on the presence or absence of substructures. Therefore, the similarity of these descriptors reflects the substructure-based similarity. We obtained the pairs of the inputs and outputs of each model in the various ED model set and compared their similarity based on substructures. The results show that the similarity of inputs and outputs (Tanimoto coefficients) of both MACCS and ECFP show a saturating curve with respect to the learning progress (**Figure 3**). This trend is correlated with the partial accuracy in **Figure 1** and the accuracy of the downstream task in **Figure 2**. These results suggest that ED models with SMILES as input recognize the substructures of chemicals that contribute to the downstream tasks in the early stage of learning.

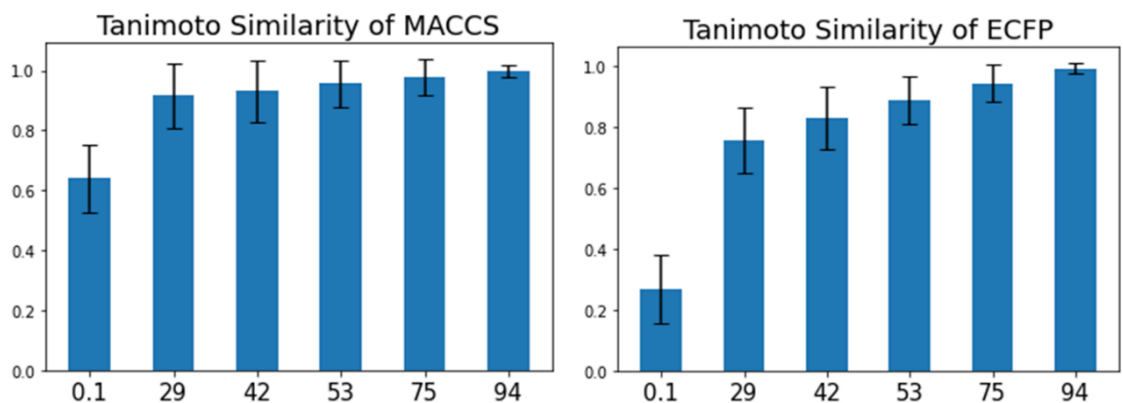

**3.4  Tendency of wrongly restored structures by ED model**

As shown in Figure 1, the progress of perfect accuracy is slower than that of partial accuracy. Taking Figures 2 and 3 into account as well, this suggests that sufficient learning is required to understand the entire chemical structure, compared with the understanding of substructures. Therefore, we worked on the evaluation of how ED models misrecognize structures when learning is insufficient. To capture the properties of misrecognized chemicals, we plotted the relationship between molecular weights of the input and the output chemicals that were wrongly restored (not matched perfectly with the inputs) by each model in the various ED model sets (Figure 4a). The results showed that molecular weights of wrongly restored chemicals were relatively large compared with the original structures. Then, we plotted the relationship between the true and predicted lengths of the strings of wrongly restored chemicals (Figure 4b). The results showed that the ED models tended to incorrectly restore the structure increasing the string length when the learning was insufficient and that this tendency was corrected as the learning progressed. This suggests that the ED model is unable to determine at which point the structure restoration is completed when learning is insufficient.

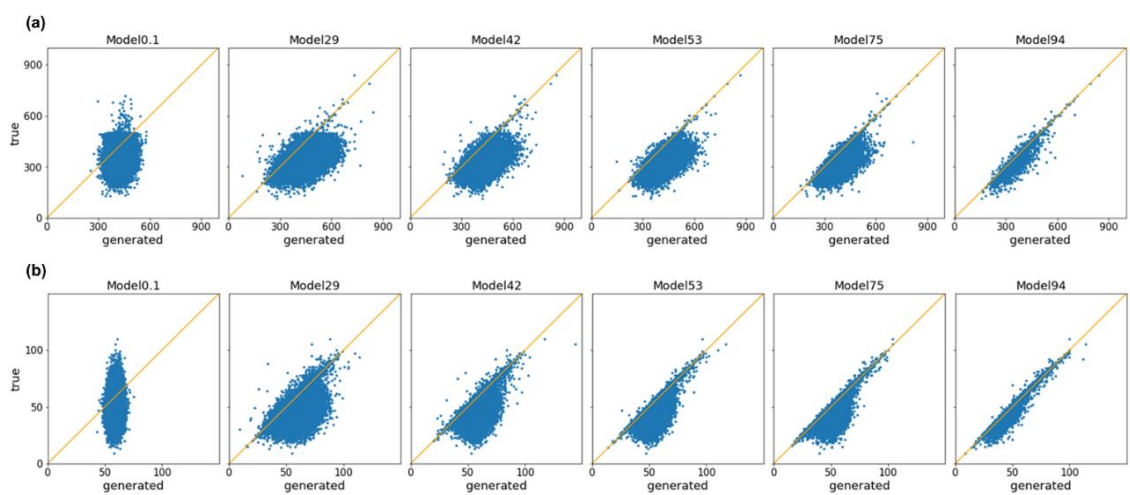

**Figure 4**. Scatterplot of (a) molecular weight and (b) SMILES length of incorrectly decoded structures. Yellow line indicates $y = x$ (value of actual SMILES is equal to that of predicted SMILES).

## 4 Conclusion

In this work, we analyzed how the structure of chemicals is recognized (acquisition of numerical representation reflecting the structure and reconstruction of the structure) during the learning progress in ED models, in which SMILES of various compounds are learned for descriptor generation.

The main contributions of this study are as follows:

- We showed that compound substructures are learned early in ED models and that existing evaluation methods based on the accuracy of downstream tasks may not be sensitive enough to evaluate the performance of ED models as descriptor generation methods.
- In addition, we showed that structure restoration is time-consuming, and in particular, insufficient learning leads to an estimation of a larger structure than the actual one. It can be inferred that determining the endpoint of the structure is a difficult task for the model.
- To our knowledge, this is the first study that connects the learning progress of SMILES representation and various chemical structures.

In this study, we employed the GRU model, inspired by the work of Winter et al., who first introduced NMT for SMILES. On the other hand, since 2017, Transformer has become the de facto standard in the field of NLP, and Transformer-based methods for generating chemical descriptors have also been developed.[40]–[44] It is an interesting future task to elucidate how Transformer-based methods recognize the chemical structure, although we focused on the GRU model in this study because of large differences between networks between Transformer and recurrent neural networks such as GRU and Long Short Term Memory (LSTM). Neural network models handling SMILES establish a field, and many models are devised not only for descriptor generation but also for de novo drug design and reaction prediction,[25], [45], [46] whereas there are still many black boxes in the relationship between the model and the chemical structure. We hope that this study will help to improve the explainability of neural network models in that field.

descriptors have also been developed[40]–[44]. It is interesting future tasks to elucidate how transformer-based methods recognize chemical structure, although we focused on the GRU model in this study due to large differences of networks between transformer and recurrent neural networks such as GRU and LSTM. Neural network models handling SMILES are establishing a field and lots of models are devised not only for descriptor generation but also for de novo drug design and reaction prediction[25], [45], [46], while there are still many black boxes in the relationship between the model and the chemical structure. We hope that this study will help to improve the explainability of neural network models in that field.

## Author Contribution

Shumpei Nemoto: Methodology, Software, Investigation, Writing – Original Draft, Visualization.
Tadahaya Mizuno: Conceptualization, Resources, Supervision, Project administration, Writing – Original Draft, Writing – Review & Editing, Funding acquisition.
Hiroyuki Kusuhara: Writing – Review & Editing

## Conflicts of Interest

The authors declare that they have no conflict of interest.

## Availability

The data sets used in this study, as well as the essential code, are available at *EDmodel* directory in https://github.com/mizuno-group/2022.

## Acknowledgement

We thank all those who contributed to the construction of the following datasets employed in the present study such as ZINC, ToxCast, and CMap[30], [31]. This work was supported by AMED under Grant Number JP22mk0101250h and the JSPS KAKENHI Grant-in-Aid for Scientific Research (C) (grant number 21K06663) from the Japan Society for the Promotion of Science.

# Supplementary Information

# Investigation of chemical structure recognition by encoder-decoder models in learning progress

Shumpei Nemoto[1]     Tadahaya Mizuno[2,†]     Hiroyuki Kusuhara[1]

[1] Laboratory of Molecular Pharmacokinetics, Graduate School of Pharmaceutical Sciences, The University of Tokyo, 7-3-1 Hongo, Bunkyo, Tokyo, Japan
[2] Laboratory of Molecular Pharmacokinetics, Graduate School of Pharmaceutical Sciences, The University of Tokyo, 7-3-1 Hongo, Bunkyo, Tokyo, Japan, tadahaya@gmail.com
†Author to whom correspondence should be addressed

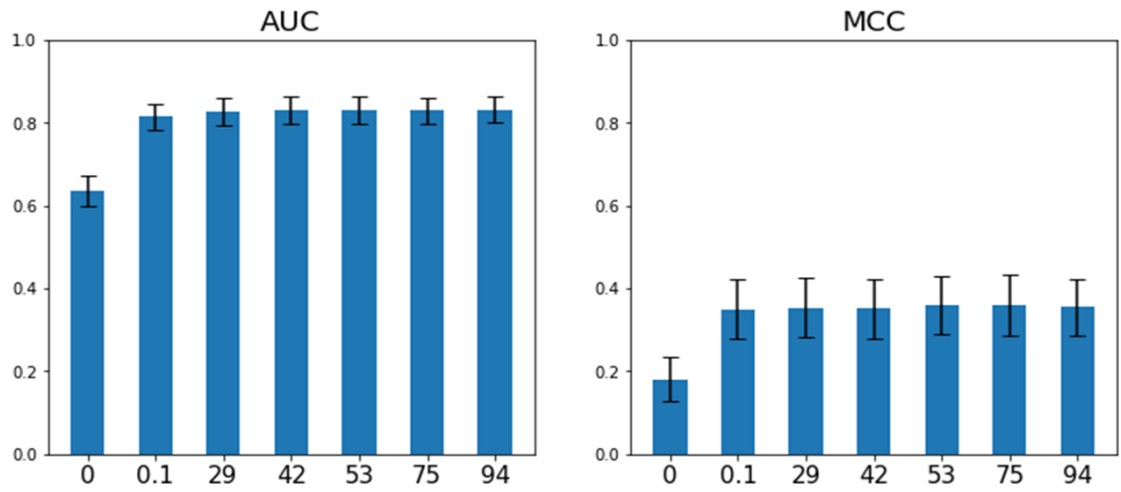

**Supplementary Figure 1. AUC and MCC of 113 assays prediction compared between perfect accuracy of Encoder-Decoder models.** Bar height and error bar indicate mean and standard.

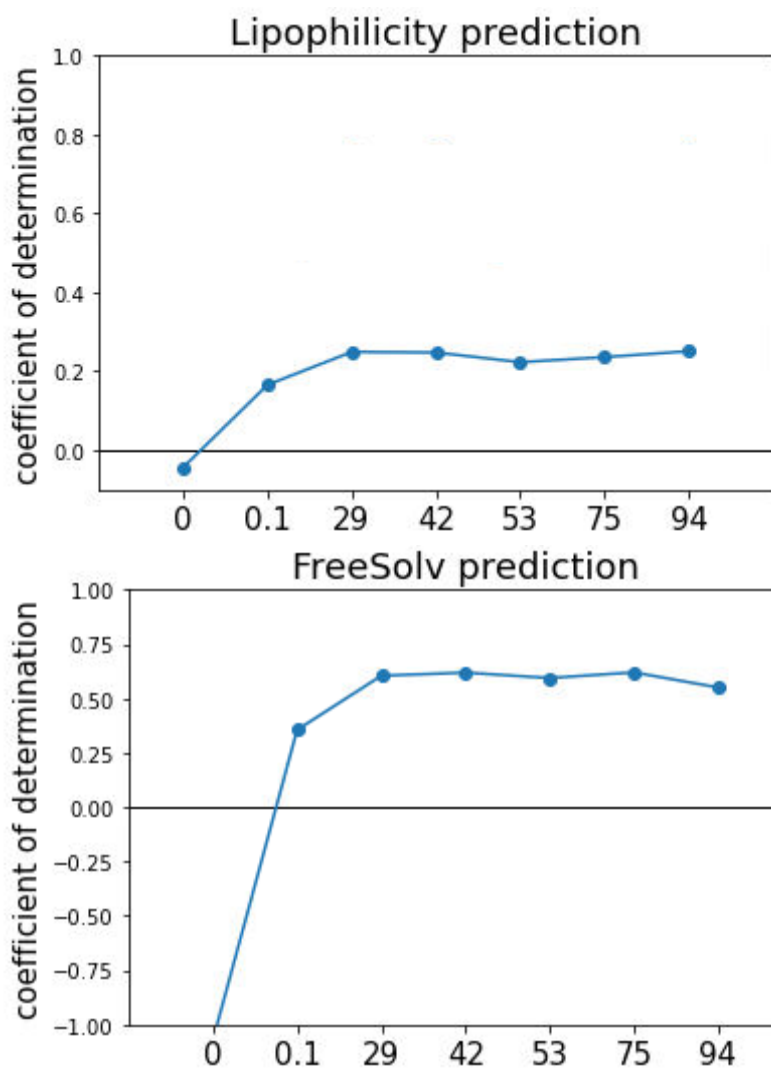

**Supplementary Figure 2. Coefficient of determination of Lipophilicity and FreeSolv prediction compared between perfect accuracy of Encoder-Decoder models.** Lipophilicity and FreeSolv data were obtained from MoleculeNet (https://moleculenet.org/). XGBoost was used as machine learning algorithm for prediction. Hyperparameters listed in **Supplementary Table 1** were optimized using Optuna for each dataset prediction with optimization index of RMSE and n_trials of 50.

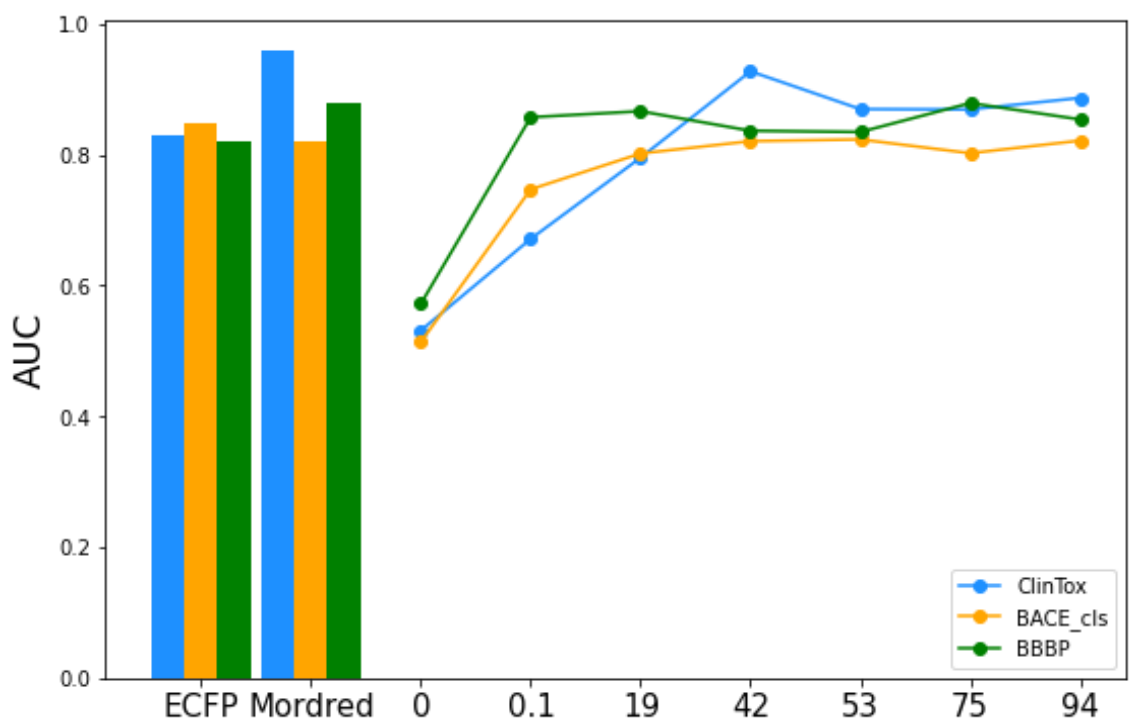

**Supplementary Figure 3. AUC of ClinTox, BACE, BBBP prediction compared between perfect accuracy of Encoder-Decoder models, ECFP and Mordred**.

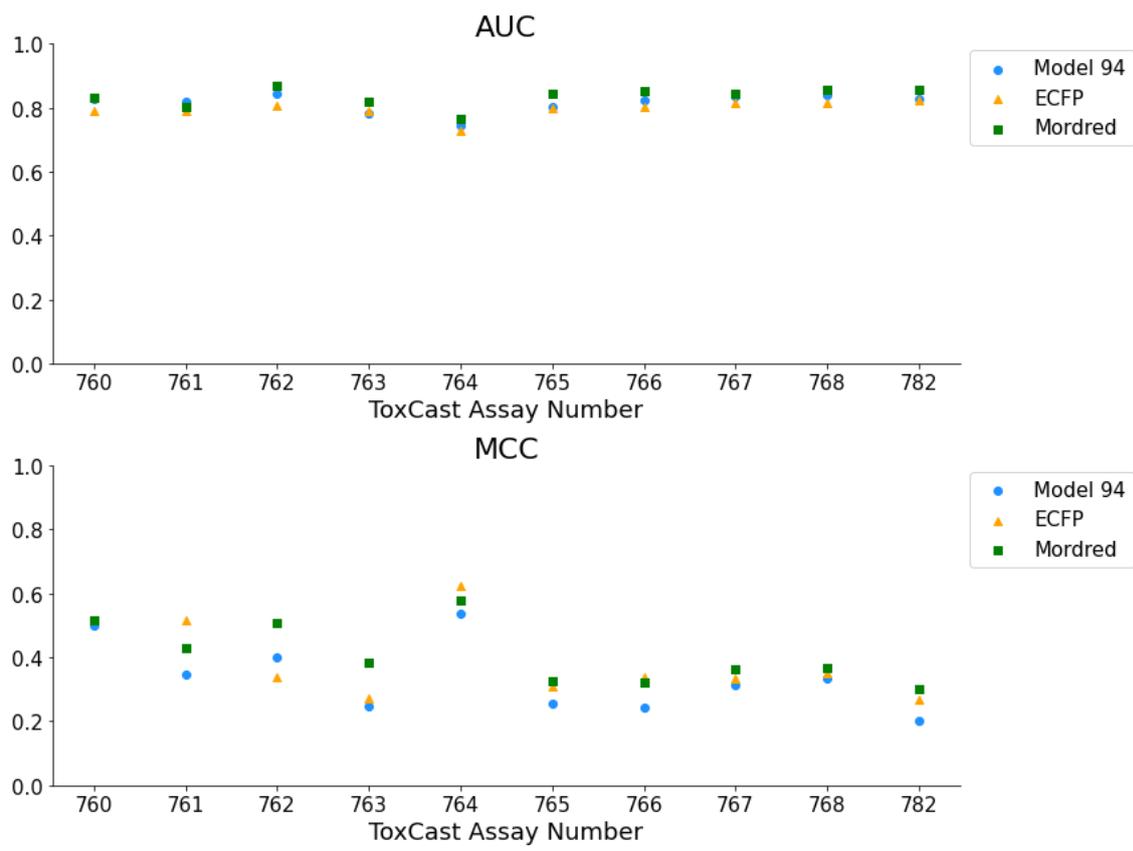

**Supplementary Figure 4. AUC and MCC of 10 assays prediction compared between Encoder-Decoder model with 94 % perfect accuracy, ECFP and Mordred.**

**Supplementary Table 1. Conditions for hyperparameter optimization in downstream task analyses** (listed in Supplementary_Table.xlsx: refer to **Availability** section).

**Supplementary Table 2. 113 HTS assays predicted by XGBoost from ToxCast.** All assays have a sample size of at least 7000 (listed in Supplementary_Table.xlsx: refer to **Availability** section).

**Supplementary Table 3. List of Tanimoto coefficients with similar compound groups and representative compounds** (listed in Supplementary_Table.xlsx: refer to **Availability** section).